\documentclass[prl,twocolumn,showpacs,superscriptaddress,preprintnumbers,amsmath,amssym]{revtex4}

\usepackage[latin1]{inputenc}
\usepackage{bm}
\usepackage{multirow,amssymb,amsbsy,amsmath}
\usepackage{graphicx}
\usepackage{verbatim}
\makeatletter
\usepackage{pifont}

\def\ket#1{\mathinner{|{#1}\rangle}}

\makeatother

\begin{document}

\title{Nonlocal memory effects in the dynamics of open quantum systems}

\author{Elsi-Mari Laine}
\affiliation{Turku Centre for Quantum Physics, Department of Physics and 
Astronomy, University of Turku, FI-20014 Turun yliopisto, Finland}

\author{Heinz-Peter Breuer}
\email{breuer@physik.uni-freiburg.de}
\affiliation{Physikalisches Institut, Universit\"at Freiburg,
Hermann-Herder-Strasse 3, D-79104 Freiburg, Germany}

\author{Jyrki Piilo}
\email{jyrki.piilo@utu.fi}
\affiliation{Turku Centre for Quantum Physics, Department of Physics and 
Astronomy, University of Turku, FI-20014 Turun yliopisto, Finland}

\author{Chuan-Feng Li}
\email{cfli@ustc.edu.cn}
\affiliation{Key Laboratory of Quantum Information, University of Science and 
Technology of China, CAS, Hefei, 230026, China}

\author{Guang-Can Guo}
\affiliation{Key Laboratory of Quantum Information, University of Science and 
Technology of China, CAS, Hefei, 230026, China}

\date{\today}

\begin{abstract}
We explore the possibility to generate nonlocal dynamical maps of an open 
quantum system through local system-environment interactions. Employing a 
generic decoherence process induced by a local interaction Hamiltonian, we show 
that initial correlations in a composite environment can lead to a nonlocal open 
system dynamics which exhibits strong memory effects although the local 
dynamics is Markovian. In a model of two entangled photons interacting with two 
dephasing environments we find a direct connection between the degree of 
memory effects and the amount of correlations in the initial environmental state. 
The results demonstrate that, contrary to conventional wisdom, enlarging an open 
system can change the dynamics from Markovian to non-Markovian.
\end{abstract}


\maketitle

Coupling a quantum mechanical systems to an external environment causes the 
system to lose information to its surroundings. Since merely all realistic quantum 
systems are open, understanding and controlling the dynamics arising from the 
presence of the environment is of central importance in present-day research 
\cite{Barreiro, Wineland}. The standard approach to the dynamics of open 
quantum systems employs the concept of a quantum Markov process which is 
given by a semigroup of completely positive dynamical maps and a corresponding 
quantum master equation with a generator in Lindblad form \cite{Lindblad, Gorini}.  
However, many quantum systems exhibit non-Markovian behavior in which there 
is a flow of information from the environment back to the open system, signifying 
the presence of quantum memory effects 
\cite{Breuer2007,Fleming,Burghardt, Piilo08,Rebentrost,Paternostro}.

For many processes occurring in nature the approximations allowing a simple 
Markovian description are not applicable. It is known, for example, that  strong 
system-environment couplings, structured and finite reservoirs, low temperatures, 
as well as the presence of large initial system-environment correlations can give 
rise to memory effects in the open system dynamics. The recognition of the 
importance of non-Markovian processes has initiated many essential steps 
towards the development of a general consistent theory of non-Markovian 
quantum dynamics  
\cite{Wolf,Lidar2009,BLP,BLP2,RHP,Kossakowski2010,Bassano1}  as well as 
achievements in the experimental detection and control of memory effects 
\cite{exp1,exp2}.

In this Letter we introduce a hitherto unexplored source for quantum memory 
effects, namely the presence of initial correlations between the subsystems
of a composite environment which interact locally with the subsystems of a 
composite open system. It is demonstrated that correlations between the 
environmental subsystems can generate a nonlocal quantum process from a 
perfectly local interaction Hamiltonian. We will show further that a nonlocal 
decoherence process can lead to non-Markovian behavior although the 
local dynamics of both subsystems is Markovian. These features are discussed
by employing two theoretical models, namely a generic decoherence model of two 
qubits interacting with correlated multimode fields, and an experimentally 
realizable model of entangled down converted photons traveling through 
birefringent media. We thus find a new, experimentally controllable source for 
memory effects in a quantum dynamical process. Besides the practical importance 
of the result in the physical realization and control of dynamical processes, it also 
reveals an unexpected feature about the nature of non-Markovian dynamics of 
composite quantum systems: Enlarging the open system can actually 
turn the dynamics from a Markovian to a non-Markovian regime.

We consider an open system $S$ consisting of two subsystems labeled by an 
index $i=1,2$, and an environment $E$ which is also composed of two 
subsystems. We assume that there are only local system-environment 
interactions, i.e., that subsystem $i$ of $S$ interacts only with its environment $i$ 
of $E$. The local interactions are described by unitaries $U_i(t)$, and 
$S$ and $E$ are supposed to be uncorrelated at the initial time. The 
open system state at time $t$ is thus given by
\begin{eqnarray*}
&&\rho_S^{12}(t)=\Phi_{12}(t)(\rho_S^{12}(0))\\
&&=\text{tr}_E\left[\big(U_1(t)\otimes U_2(t)\big)\rho_S^{12}(0)\otimes 
\rho_E^{12}(0) \big(U_1^{\dagger}(t)\otimes U_2^{\dagger}(t)\big)\right],
\end{eqnarray*}
where $\Phi_{12}(t)$ represents the quantum dynamical map describing the
time evolution of $S$. If the two environments are initially uncorrelated, 
$ \rho_E^{12}(0)=\rho_E^1(0)\otimes\rho_E^2(0)$, this map factorizes and the 
dynamics of $S$ is given by a product of local maps, 
$\Phi_{12}(t)=\Phi_{1}(t)\otimes\Phi_{2}(t)$. 
However, when $\rho_E^{12}(0)$ exhibits correlations $\Phi_{12}(t)$ does not in 
general factorize and the environmental correlations may give rise to a nonlocal 
process even when the interaction Hamiltonian is purely local. For a local map the 
dynamical properties of the subsystems completely determine the global system 
dynamics, but when the map is nonlocal the global system can exhibit features 
which are not present in the dynamics of the individual subsystems. Here we 
explore especially quantum memory effects arising from a nonlocal dynamics.

We consider a dephasing map for two qubits of the general form 
\begin{equation} \label{rho12}
 \rho_S^{12}(t)=\left(\begin{array}{cccc}
 |a|^2 & a b^* \kappa_2(t) & ac^* \kappa_1(t)& ad^* \kappa_{12}(t) \\
 b a^* \kappa_2^*(t) & |b|^2 & b c^*\Lambda_{12}(t) & b d^* \kappa_1(t)\\
 c a^* \kappa_1^*(t) & c b^* \Lambda_{12}^*(t) & |c|^2 & c d^* \kappa_2(t)\\
 d a^* \kappa_{12}^*(t) & d b^* \kappa_1^*(t) & d c^* \kappa_2^*(t) & |d|^2
\end{array}\right),
\end{equation}
where the initial state of the two qubit system is a pure state given by
\begin{equation}\label{init}
 \ket{\psi_{12}}=a \ket{00}+b\ket{01}+c\ket{10}+d\ket{11}.
\end{equation}
The corresponding dynamics for subsystems $1$ and $2$ are given by 
$\rho_S^1(t)=\text{tr}_2\left[\rho_S^{12}(t)\right]$ and 
$\rho_S^2(t)=\text{tr}_1\left[\rho_S^{12}(t)\right]$. 
The states $\rho_S^1(t)$ and $\rho_S^2(t)$ are fully determined by the functions 
$\kappa_1(t)$ and $\kappa_2(t)$, depending neither on 
$\kappa_{12}(t)$ nor on $\Lambda_{12}(t)$.
The interaction Hamiltonian is assumed to be local, i.e., we have
\begin{equation} \label{eq1}
 H_{\rm int}(t) = \chi_1(t) H_1 + \chi_2(t) H_2,
\end{equation}
where the function $ \chi_i(t)$ is $1$ for $ t_i^{\rm s} \leq t \leq t_i^{\rm f}$ and zero 
otherwise. Here, $t_i^{\rm s}$ and $t_i^{\rm f}$ denote the times the interaction is 
switched on and switched off in system $i$, respectively. Since the local 
Hamiltonians $H_i$ commute, the time evolution of the total system is given by 
$|\Psi(t)\rangle = \exp\left[-i\int_0^t dt' H_{\rm int} (t') \right] |\Psi(0)\rangle$. We will 
further denote the local interaction times as $t_i(t) = \int_0^t \chi_{i} (t')dt'$ and for 
convenience we will not explicitly write the time dependence of $t_i$.

Before turning to the details of the physical systems under study, let us briefly 
discuss the concept of memory effects. Memory effects are quantified in 
\cite{BLP} by employing the trace 
distance $D(\rho_A, \rho_B) = \frac{1}{2} \textrm{tr}|\rho_A-\rho_B|$ between two 
quantum states $\rho_A$ and $\rho_B$. This quantity can be interpreted as a 
measure for the distinguishability of the two states \cite
{Hellstrom,Holevo,Hayashi}. In view of this interpretation the 
characteristic feature of a non-Markovian quantum process is the increase of the 
distinguishability, i.e.~a reversed flow of information from the environment back to 
the open system. Through this recycling of information the earlier states of the 
open system influence its later states \cite{Piilo08}, which expresses the 
emergence of memory effects in the open system dynamics. The measure for 
non-Markovianity is written as
\begin{equation}\label{eq4}
 \mathcal{N}(\Phi)=\max_{\rho_{A,B}(0)}\int_{\sigma>0}dt \,
 \sigma(t,\rho_{ A,B}(0)),
\end{equation}
where $\sigma(t,\rho_{A,B}(0))=\frac{d}{dt}D(\rho_A(t),\rho_B(t))$. Here, the 
time integration is extended over all subintervals of time in which the rate of 
change of the trace distance $\sigma$ is positive, and the maximum is taken over 
all pairs of initial states. The quantity in Eq. \eqref{eq4} thus measures the 
maximal total amount of information which flows from the environment back to 
the open system over the whole time evolution.

First, we study a generic model of two qubits interacting with correlated multimode 
fields. The local interaction Hamiltonians of Eq.~(\ref{eq1}) are $H_{i} =\sum_k
\sigma_z^i(g_k b_k^{i\dagger}+g_k^*b_k^i)$. We assume that the 
interaction strengths in both systems are identical, $g_k^1=g_k^2$. The local 
time evolution of the systems is then given by the unitary
\begin{equation}
 U_{i}(t) =
 \exp\Big\{\sigma_z^{i}\sum_k\big(b_k^{i \dagger}\xi_k(t_{i})-b_k^i\xi_k^*(t_{i})
 \big)\Big\},
 \label{UA}
\end{equation}
where $\xi_k(t_{i})=g_k(1-e^{i\omega_k t_{i}})/\omega_k$.
The local unitary of Eq.~(\ref{UA}) acts in the following way:
\begin{eqnarray*}
U_{i}(t)\ket{0}\otimes\ket{\eta}&=&\ket{0}\otimes \bigotimes_k D(-\xi_k(t_{i})) \ket
{\eta},\nonumber\\
U_{i}(t)\ket{1}\otimes\ket{\eta}&=&\ket{1}\otimes \bigotimes_k D(\xi_k(t_{i})) \ket
{\eta},\nonumber
\end{eqnarray*}
where $D(\xi_k)$ is the displacement operator for the $k$th mode. Let us take as 
the initial state $\ket{\Psi(0)}=\ket{\psi_{12}}\otimes \ket{\eta_{12}}$, where $\ket
{\psi_{12}}$ is given by Eq.~(\ref{init}) and $\ket{\eta_{12}}=\bigotimes_k\ket{\eta_
{12}^k}$. The decoherence process is then given by Eq.~(\ref{rho12}), where
$\kappa_1(t)=\langle\eta_{12}^{10}|\eta_{12}^{00}\rangle$,
$\kappa_2(t)=\langle\eta_{12}^{01}|\eta_{12}^{00}\rangle$,
$\kappa_{12}(t)=\langle\eta_{12}^{11}|\eta_{12}^{00}\rangle$, and
$\Lambda_{12}(t)=\langle\eta_{12}^{10}|\eta_{12}^{01}\rangle$
with
\begin{equation*}
\ket{\eta_{12}^{nm}(t)}=\bigotimes_k \left[D((-1)^{n+1}\xi_k(t_1))\otimes 
D((-1)^{m+1}\xi_k(t_2))\right]\ket{\eta_{12}}.
\end{equation*}
After some algebra, one finds
\begin{eqnarray*}
\kappa_1(t)&=&\prod_k\chi_k(-2 \xi_k(t_1),0),\\ 
\kappa_2(t)&=&\prod_k\chi_k(0,-2 \xi_k(t_2)),\\
\kappa_{12}(t)&=&\prod_k\chi_k(-2 \xi_k(t_1),-2 \xi_k(t_2)),\\
\Lambda_{12}(t)&=&\prod_k\chi_k(-2 \xi_k(t_1),+2 \xi_k(t_2)), 
\end{eqnarray*}
where $\chi_k(x,y)$ is the characteristic function of $\ket{\eta_{12}^k}$.

\begin{figure}[h]
\centering
\includegraphics[width=0.45\textwidth]{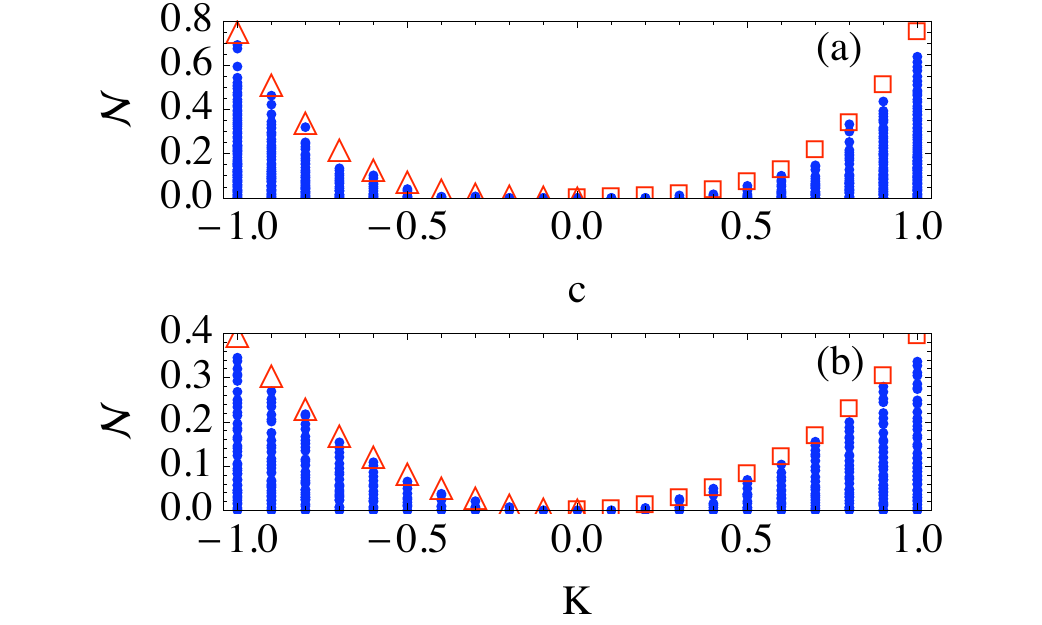}
\caption{\label{Fig:1}(color online) Maximization over the pairs of initial states. The 
(blue) dots represent the increase of the trace distance for 1000 
randomly drawn initial pairs of states. (a) Two qubits interacting with multimode 
fields. The measure for non-Markovianity for different 
values of $c$ and for $\alpha=1$. The (red) triangles represent the measure 
for the maximizing pair $\frac{1}{\sqrt{2}}(\ket{00}\pm\ket{11})
$, and the (red) squares for $\frac{1}{\sqrt
{2}}(\ket{01}\pm\ket{10})$. (b) Two photons moving through quartz plates. The 
measure for non-Markovianity for different 
values of the correlation coefficient $K$, and a fixed $C_{11}^{1/2} T \Delta 
n=1$. The (red) triangles represent the measure 
for the maximizing pair $\frac{1}{\sqrt{2}}(\ket{HH}\pm\ket{VV})
$, and the (red) squares for $\frac{1}{\sqrt
{2}}(\ket{HV}\pm\ket{VH})$.}
\end{figure}

\begin{figure}[h]
\centering
\includegraphics[width=0.35\textwidth]{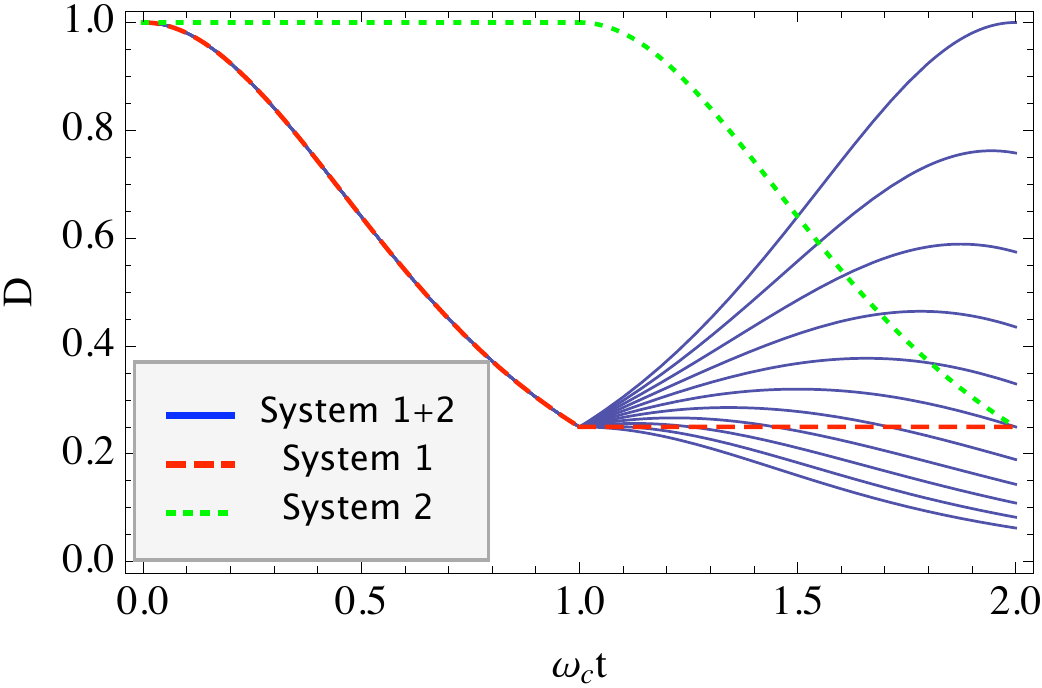}
\caption{\label{Fig:2}(color online) The trace distance dynamics for the two qubits 
interacting with correlated multimode fields. We take $\alpha=1$, 
$t_1^{\rm s}=0$, $t_1^{\rm f}=1=t_2^{\rm s}$ and $t_2^{\rm f}=2$ in units of $
\omega_c^{-1}$. The solid (blue) lines represent the trace distance with different 
values of $c$ for the global dynamics of the two qubits for the maximizing pair of 
initial states. The dashed (red) line and the dotted (green) line give the trace 
distance evolution for the initial states $1/\sqrt{2}(\ket{0}\pm\ket{1})$ in systems 
$1$ and $2$, respectively.}
\end{figure}

Let us consider a two-mode Gaussian state with the characteristic function 
$\chi_k(x,y)=\chi_k(\lambda_1,\lambda_2,\lambda_3,\lambda_4)
=\exp(-\frac{1}{2}\vec{\lambda}^T\boldsymbol{\sigma}\vec{\lambda})$,
where $\lambda_1=\Re[x]$, $\lambda_2=\Im[x]$, $\lambda_3=\Re[y]$, 
$\lambda_4=\Im[y]$ and
\begin{displaymath}
\boldsymbol{\sigma}=\left(\begin{array}{cc}
A&C\\
C^T&B
\end{array}\right), \nonumber
\end{displaymath}
is the covariance matrix of the state. Let us take $A=B=\mathbb{I}$ and 
$C=c\mathbb{I}$. Now the state is uncorrelated iff $c=0$. We can write
\begin{equation*}
\chi_k(x,y)=\exp\left[-\frac{1}{2}\left\{|x|^2+|y|^2+c(xy^*+x^*y)\right\}\right].
\end{equation*}
For $c=-1$ we get $\chi_k(x,y)=\exp\left[-\frac{1}{2}|x-y|^2\right]$ and 
$\kappa_{12}(t)=\exp\left[-2\sum_k|\xi_k(t_1)-\xi_k(t_2)|^2\right]$. Performing the 
continuum limit we obtain
\begin{equation}
\kappa_{12}(t)=\exp\left\{-4\int_0^\infty d\omega
J(\omega)\frac{1-\cos\left[\omega|t_1(t)-t_2(t)|\right]}{\omega^2}\right\},\nonumber
\end{equation}
where $J(\omega)$ is the spectral density of the reservoir. For an ohmic spectral 
density $J(\omega)=\alpha \omega \exp(-\omega/\omega_c)$ with coupling 
constant $\alpha$ and frequency cutoff $\omega_c$ we have
\begin{eqnarray*}
\kappa_1(t)&=&\left(1+\omega_c^2 t_1^2(t)\right)^{-2 \alpha},\\
\kappa_2(t)&=&\left(1+\omega_c^2 t_2^2(t)\right)^{-2 \alpha},\\
\kappa_{12}(t)&=&\left(1+\omega_c^2 |t_1(t)-t_2(t)|^2\right)^{-2 \alpha},\\
\Lambda_{12}(t)&=&\kappa_1^2(t)\kappa_2^2(t)/\kappa_{12}(t).
\end{eqnarray*}
The maximization over the initial states in Eq.~(\ref{eq4}) for different values of $c$ 
is presented in Fig.~\ref{Fig:1}(a). The trace distance dynamics of the subsystems 
$1$ and $2$ as well as the global dynamics are presented in Fig.~\ref{Fig:2}. We 
clearly see that the trace distance in the subsystems $1$ and $2$ continuously 
decreases, but for the total system the trace distance does indeed increase: We 
obtain a dynamics which is locally Markovian but globally exhibits nonlocal 
memory effects.

As our second example we examine an experimentally realizable model of a pair 
of entangled photons subjected to local birefringent 
environments~\cite{Kwiat00,Xu10}. The photon pair is created in a spontaneous 
parametric down-conversion process after which the photons separate,
traveling along different arms $i=1,2$, and move through different quartz plates.
When a photon enters a quartz plate a local interaction between the polarization 
degrees of freedom (forming the open system) and the frequency degrees 
of freedom (forming the environment) is switched on. The Hamiltonian describing 
the local interaction in Eq.~(\ref{eq1}) of a photon induced by the corresponding 
quartz plate is given by
\begin{equation}
 H_i = -\int d\omega_i \, \omega_i \Big[ n_V |V\rangle\langle V| +
 n_H |H\rangle\langle H| \Big] 
 \otimes |\omega_i\rangle\langle\omega_i|, \nonumber
\end{equation}
where $|\lambda \rangle \otimes |\omega_i\rangle$ denotes the state of a 
photon in arm $i$ with polarization $\lambda = H,V$ (horizontal or vertical) and 
frequency $\omega_i$. The refraction index of the polarization state $\lambda$ is 
denoted by $n_{\lambda}$. The total initial state is given by $\ket{\Psi(0)} = \ket{\psi_{12}} \otimes \int d\omega_1 d\omega_2 \,
 g(\omega_1,\omega_2) \ket{\omega_1,\omega_2}$, where $\ket{\psi_{12}}=a \ket{HH}+b\ket{HV}+c\ket{VH}+d\ket{VV}$.
Initially the environment formed by the mode degrees of freedom is thus in a 
correlated state with $g(\omega_1,\omega_2)$ denoting the amplitude of finding a 
photon with frequency $\omega_1$ in arm $1$ and a photon with frequency $
\omega_2$ in arm $2$. The corresponding joint probability distribution will be 
denoted by $P(\omega_1,\omega_2)=|g(\omega_1,\omega_2)|^2$.

The state of the open system (polarization states) at time $t$ is of the form of 
Eq.~(\ref{rho12}) with the functions $\kappa_1(t)=G(\Delta n t_1,0)$, 
$\kappa_2(t)=G(0,\Delta n t_2)$, 
$\kappa_{12}(t)=G(\Delta n t_1,\Delta n t_2)$, and 
$\Lambda_{12}(t)=G(\Delta n t_1,-\Delta n t_2)$,
where 
\[ G(\tau_1,\tau_2)=\int d\omega_1 d\omega_2 P(\omega_1,\omega_2) 
e^{-i(\omega_1\tau_1+\omega_2\tau_2)}
\]
is the Fourier transform of $P(\omega_1,\omega_2)$ and 
$\Delta n = n_V - n_H$ is the birefringence. Note that although the Hamiltonian of 
(\ref{eq1}) is a sum of local interaction terms, the corresponding dynamical map 
$\Phi_{12}(t)$ is a product of local dynamical maps if and only if 
$\kappa_{12}(t)=\kappa_1(t)\kappa_2(t)$ and 
$\Lambda_{12}(t)= \kappa_1(t)\kappa_2^*(t)$. This is the case only when the joint 
frequency distribution $P(\omega_1,\omega_2)$ factorizes, i.e., when the 
frequencies $\omega_1$ and $\omega_2$ are uncorrelated.

To characterize the correlations in the initial environmental state we introduce the
covariance matrix $C=(C_{ij})$ with elements $C_{ij}=\langle\omega_i\omega_j
\rangle-\langle\omega_i\rangle\langle\omega_j\rangle$. We will assume that both 
the means and the variances of $\omega_1$ and $\omega_2$ are equal, i.e., $
\langle \omega_1 \rangle = \langle \omega_2 \rangle =\omega_0/2$ and $C_{11} = 
C_{22} = \langle\omega^2_i\rangle-\langle\omega_i\rangle^2$. To quantify the 
frequency correlations we use the correlation coefficient $K=C_{12}/\sqrt{C_{11}
C_{22}}=C_{12}/C_{11}$. We have $|K|\leq1$,where the equality sign holds, i.e., 
$K=\pm1$, if and only if $\omega_1$ and $\omega_2$ are linearly related.

Let us take a Gaussian frequency distribution
\begin{equation} \label{GAUSS}
 P(\omega_1,\omega_2) = \frac{1}{2\pi\sqrt{\textrm{det}C}}
 e^{-\frac{1}{2}
 \left(\vec{\omega}-\vec{\langle \omega \rangle }\right)^TC^{-1}
 \left(\vec{\omega}-\vec{\langle \omega \rangle }\right)},
\end{equation}
where $\vec{\omega}=(\omega_1,\omega_2)^T$ and $\vec{\langle \omega 
\rangle }=(\langle \omega_1 \rangle,\langle \omega_2 \rangle)^T$. One can easily 
find the Fourier transform of this distribution,
\begin{equation} \label{Gtau1tau2}
 G(\tau_1,\tau_2)
 = e^{i\omega_0(\tau_1+\tau_2)/2 - C_{11}\left(\tau_1^2 + \tau_2^2
 + 2K\tau_1\tau_2\right)/2} .
\end{equation}

\begin{figure}[]
\centering
\includegraphics[width=0.4\textwidth]{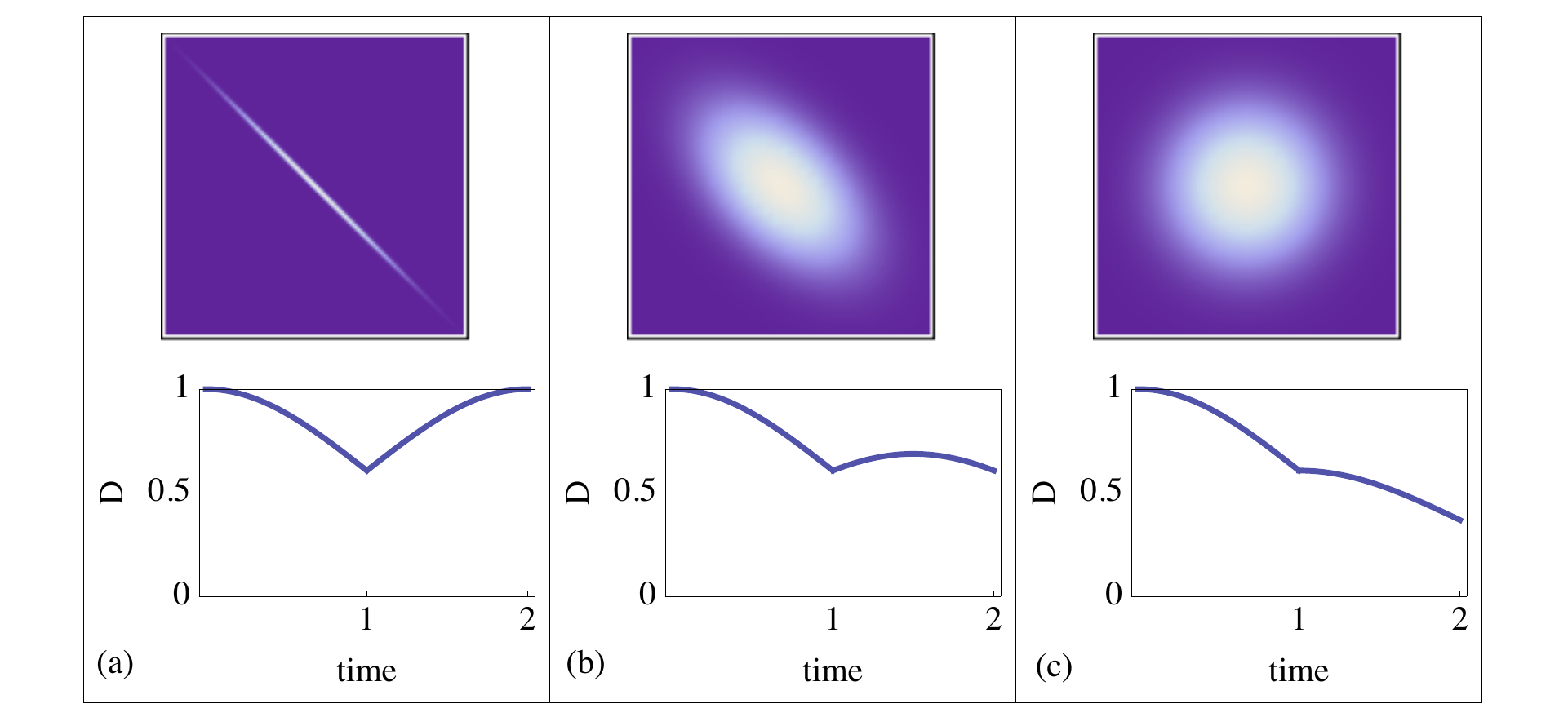}
\caption{\label{Fig:3}(color online) The connection between the frequency distribution 
$P(\omega_1,\omega_2)$ (top) and the dynamics of the trace distance (bottom) 
for different values of $K$. (a) $K=-1$, (b) $K=-0.5$, (c) $K=0$. The unit of time is $\sqrt{C_{11}} T \Delta n$.}
\end{figure}

We assume for simplicity that the total interaction times for both photons are 
equal, denoting it by $T=t_1^{\rm f} - t_1^{\rm s} =  t_2^{\rm f} - t_2^{\rm s}$,
 and that the quartz plates are mounted one after 
another, i.e., $t_1^{\rm f}=t_2^{\rm s}$. We can then derive an analytic 
expression for the measure \eqref{eq4} of the non-Markovianity of the process. 
The maximization over the pair of initial states in Eq. \eqref{eq4} is illustrated in Fig.~\ref{Fig:1}(b).
Using Eqs. \eqref{rho12} and \eqref{Gtau1tau2} we obtain the time dependence of the trace distance for the maximizing initial pairs,
\begin{equation}
 D(t)= \exp\left[
 -\frac{\Delta n^2}{2}C_{11} \left( t_1^2 + t_2^2 - 2|K|t_1t_2 \right) \right].
 \nonumber
\end{equation}
During the interaction of the photon in arm $1$ the trace distance first decreases 
from the initial value $1$ to the value $D_1=\exp\left[-\frac{\Delta n^2}{2}C_{11} T^2 \right]$.
The subsequent interaction of the photon in arm $2$ depends on the function 
$f(t_2) = t_2^2 - 2|K|Tt_2$ for $t_2 \in [0,T]$. This function decreases 
monotonically in the interval $[0,|K|T]$ from the value $f(0)=0$ to the value 
$f(|K|T)=-(KT)^2$, which means that the trace distance increases over this 
interval to the value $D_2 = \exp\left[
 -\frac{\Delta n^2}{2}C_{11} \left( T^2 - (KT)^2 \right) \right]$. It follows that the
non-Markovianity measure is given by
\begin{equation} \label{eq6} 
 {\mathcal{N}}  = D_2 - D_1
 = e^{-\frac{1}{2}C_{11}(\Delta n T)^2}
 \left[ e^{\frac{1}{2}C_{11}(\Delta n T)^2K^2} - 1 \right].
\end{equation}
This equation establishes a direct connection between the measure for 
non-Markovianity and the degree of correlations in the initial environmental state 
as quantified by the correlation coefficient $K$. We also see that the process is 
Markovian if and only if $K=0$. The relation \eqref{eq6} is further illustrated in
Fig. \ref{Fig:3}, where we have plotted the frequency distribution 
$P(\omega_1,\omega_2)$ and the dynamics of the trace distance for three different values of the correlation coefficient, $K=0.0$, $-0.5$, $-1.0$ (anticorrelation). One clearly observes that when the frequencies $\omega_1$ and $\omega_2$ become more anticorrelated the dynamics becomes more non-Markovian. In general, we conclude that the reduced dynamics of the two-photon polarization state
is non-Markovian whenever the frequency distribution 
$P(\omega_1,\omega_2)$ exhibits correlations. This behavior occurs globally, i.e., 
when we study the dynamics of the composite state of both photons. However, if 
one observes the local dynamics of either of the photons the 
process is always Markovian.  

Summarizing, we have found a new source for memory effects in the dynamics of 
open quantum systems. We studied a generic dephasing model as well as a theoretical scheme which is experimentally 
realizable with current technology. The initial correlations within the environment induce a nonlocal map which gives rise to 
memory effects in the open system dynamics. Locally, each subsystem follows 
Markovian dynamics, but globally they are exposed to memory effects even 
though the interaction Hamiltonian is local. Since for 
classical stochastic processes a non-Markovian process can be embedded in a 
Markovian one by a suitable enlargement of the number of relevant variables, the 
general view has been that enlarging the open quantum system under study tunes 
the dynamics towards a Markovian behavior. This can be done for certain
non-Markovian quantum processes as well \cite{Hughes1, Plenio10, Siegle, 
Bassano2}, but we see that also the exactly opposite behavior can occur, i.e.,  
enlarging the subsystem can bring the dynamics from a Markovian to a 
non-Markovian regime.

\acknowledgments
This work was supported by the National Fundamental Research Program, 
National Natural Science Foundation of China (Grant Nos.~60921091, 10874162 
and 10734060), the Academy of Finland (Project 259827), the Jenny and Antti 
Wihuri Foundation, the Graduate School of Modern Optics and Photonics, and the 
German Academic Exchange Service (DAAD). EML, HPB, and JP thank the USTC 
group for the hospitality during a research visit at the Key Laboratory of Quantum 
Information.


\begin{thebibliography}{xx}

\bibitem{Barreiro}
J. T. Barreiro,  M. M\"{u}ller,  P. Schindler,  D. Nigg,  T. Monz,  M. Chwalla,  M. 
Hennrich,  C. F. Roos,  P. Zoller, and R. Blatt, Nature {\textbf{470}}, 486 (2011).

\bibitem{Wineland}
C. J. Myatt, B. E. King, Q. A. Turchette, C. A. Sackett, D. Kielpinski, W. M. Itano, C. 
Monroe, and D. J. Wineland, Nature {\textbf{403}}, 269 (2000).

\bibitem{Lindblad}
G. Lindblad,
Commun. Math. Phys. {\textbf{48}}, 119 (1976).

\bibitem{Gorini}
V. Gorini,A. Kossakowski, and E. C. G. Sudarshan,
J. Math. Phys. {\textbf{17}}, 821 (1976).

\bibitem{Breuer2007}
H.-P Breuer and F. Petruccione,
{\textit{The Theory of Open Quantum Systems}}
(Oxford University Press, Oxford,  2007).

\bibitem{Fleming}
H. Lee, Y.-C. Cheng, and G. R. Fleming,
Science  {\textbf{316}}, 1462 (2007).

\bibitem{Burghardt}
 L. S. Cederbaum, E. Gindensperger, and I. Burghardt,
Phys. Rev. Lett.  {\textbf{94}}, 113003 (2005).

\bibitem{Piilo08}
J. Piilo, S. Maniscalco, K. H\"ark\"{o}nen, and K.-A. Suominen,
Phys. Rev. Lett. {\textbf{100}}, 180402 (2008);  
J. Piilo, K. H\"ark\"{o}nen, S. Maniscalco, and K.-A. Suominen,
Phys. Rev. A  {\textbf{79}}, 062112 (2009).

\bibitem{Rebentrost}
P. Rebentrost and A. Aspuru-Guzik,
J. Chem. Phys.  {\textbf{134}}, 101103 (2011).

\bibitem{Paternostro}
T. J. G. Apollaro, C. Di Franco, F. Plastina, and M. Paternostro,
Phys. Rev. A  {\textbf{83}}, 032103 (2011).

\bibitem{Wolf}
M. M. Wolf, J. Eisert, T.S. Cubitt, and J.I. Cirac,
Phys. Rev. Lett. {\textbf{101}}, 150402 (2008).

\bibitem{Lidar2009}
A. Shabani and D. A. Lidar, Phys. Rev. Lett. \textbf{102}, 100402 (2009).

\bibitem{BLP}
H.-P. Breuer, E.-M. Laine, and J. Piilo,
Phys. Rev. Lett. {\textbf{103}}, 210401 (2009).

\bibitem{BLP2}
E.-M. Laine, J. Piilo, and H.-P. Breuer,
Phys. Rev. A {\textbf{81}}, 062115 (2010).

\bibitem{RHP}
A. Rivas, S. F. Huelga, and  M. B. Plenio,
Phys. Rev. Lett. {\textbf{105}}, 050403 (2010).

\bibitem{Kossakowski2010}
D. Chru\'sci\'nski and A. Kossakowski,
Phys. Rev. Lett. \textbf{104}, 070406 (2010).


\bibitem{Bassano1}
B. Vacchini, A. Smirne, E.-M. Laine, J. Piilo, and H.-P. Breuer,
New J. Phys. {\textbf{13}}, 093004 (2011).

\bibitem{exp1}
J.-S. Tang, C.-F. Li, Y.-L. Li, X.-B. Zou, G.-C. Guo, H.-P. Breuer, E.-M. Laine, and J. 
Piilo, EPL {\textbf{97}}, 10002 (2012).

\bibitem{exp2}
B.-H. Liu, L. Li, Y.-F. Huang, C.-F. Li, G.-C. Guo, E.-M. Laine, H.-P. Breuer, and J. 
Piilo, Nature Physics {\textbf{7}}, 931-934 (2011).

\bibitem{Hellstrom}
C. W. Helstrom,
{\textit{Quantum Detection and Estimation Theory}}
(Academic Press, New York,  1976).

\bibitem{Holevo}
 A. S. Holevo,
Trans. Moscow Math. Soc. {\textbf{26}}, 133 (1972).

\bibitem{Hayashi}
M. Hayashi,
{\textit{Quantum Information}}
(Springer-Verlag, Berlin, 2006).

\bibitem{Kwiat00}
P. G. Kwiat, A. J. Berglund, J. B. Altepeter, and A. G. White, Science {\textbf 
{290}}, 498 (2000).

\bibitem{Xu10}
J.-S. Xu, X.-Y. Xu, C.-F. Li,  C.-J. Zhang,  X.-B.  Zou, G.-C. Guo, Nature Commun. 
\textbf{1}, 7 (2010).

\bibitem{Hughes1}
K. H. Hughes, C. D. Christ, and I. Burghardt,
J. Chem. Phys. {\textbf{131}}, 024109 (2009).

\bibitem{Plenio10}
J. Prior, A. W. Chin, S. F. Huelga, and M. B. Plenio,
Phys. Rev. Lett. {\textbf{105}}, 050404 (2010).

 \bibitem{Siegle}
P. Siegle, I. Goychuk, P. Talkner, and P. H\"anggi,
Phys. Rev. E {\textbf{81}}, 011136 (2010).

\bibitem{Bassano2}
R. Martinazzo, B. Vacchini, K. H. Hughes, and I. Burghardt
J. Chem. Phys. {\textbf{134}}, 011101 (2011).

\end{thebibliography}
\end{document}